\documentclass{jfm}

\usepackage{graphicx}
\usepackage{epsfig}
\usepackage{newtxtext}
\usepackage{newtxmath}
\usepackage{natbib}
\usepackage{hyperref}
\usepackage{upgreek}
\usepackage{bm}
\hypersetup{
    colorlinks = true,
    urlcolor   = blue,
    citecolor  = black,
}

\newcommand{\RomanNumeralCaps}[1]
\linenumbers

\title{Microtomographic PIV measurements of viscoelastic instabilities in a 3D micro-contraction}

\author{Daniel W. Carlson,
Amy Q.\ Shen,
 \and Simon J.\ Haward \corresp{\email{simon.haward@oist.jp}}}

\affiliation{Micro/Bio/Nanofluidics Unit, Okinawa Institute of Science and Technology Graduate University, Onna, Okinawa 904-0495, Japan.}

\begin{document}
\maketitle

\begin{abstract}
Viscoelastic flow through an abrupt planar contraction geometry above a certain Weissenberg number ($Wi$) is well known to become unstable upstream of the contraction plane via a central jet separating from the walls and forming vortices in the salient corners. Here, for the first time we consider three-dimensional (3D) viscoelastic contraction flows in a microfabricated glass square-square contraction geometry. We employ state-of-the-art microtomographic particle image velocimetry to produce time-resolved and volumetric quantification of the 3D viscoelastic instabilities arising in a dilute polymer solution driven through the geometry over a wide range of $Wi$ but at negligible Reynolds number. Based on our observations, we describe new insights into the growth, propagation, and transient dynamics of an elastic vortex formed upstream of the 3D micro-contraction due to flow jetting towards the contraction. At low $Wi$ we observe vortex growth for increasing $Wi$, followed by a previously unreported vortex growth plateau region. In the plateau region, the vortex circulates around the jet with a period that decreases with $Wi$ but an amplitude that is independent of $Wi$. In addition, we report new out-of-plane asymmetric jetting behaviour with a phase-wise dependence on $Wi$. Finally, we resolve the rate-of-strain tensor $\pmb{D}$ and ascribe local gradients in $\pmb{D}$ as the underlying driver of circulation via strain-hardening of the fluid in the wake of the jet.
\end{abstract}


\section{Introduction}
\label{sec:intro}

Entry flow has historically received attention as a canonical case for non-Newtonian fluid dynamics (\cite{boger1987viscoelastic, white1987review}), and as a benchmark for developing computational models capable of studying highly elastic flows (\cite{ afonso2011dynamics,pimenta2017stabilization}). Under negligible inertia (i.e., Reynolds numbers $Re \ll 1$), for Weissenberg numbers ($Wi =\lambda \dot{\gamma}$, where $\lambda$ is the fluid relaxation time and $\dot{\gamma}$ the shear rate) beyond a critical value $Wi_c \approx 0.5$, pipe flow moving towards a contraction becomes sufficiently elastic that it separates from the upstream walls, forming a central `jet' that enters the constriction and vortices around the mouth of the constriction (\cite{mckinley1991nonlinear, rothstein1999extensional}). Initially the corner vortices are static in their placement as $Wi$ is increased, but eventually grow in size until a Hopf bifurcation characterized by a periodic fluctuation of the vortex separation point occurs. For $Wi > Wi_p \gg Wi_c$, the vortices become increasingly unsteady for increasing $Wi$ (\cite{mckinley1991nonlinear, rothstein1999extensional, rothstein2001axisymmetric}), and may lead to a period-doubling route to chaos (\cite{mckinley1991nonlinear}). The onset and subsequent dynamics of this elastic flow instability are highly sensitive to the contraction geometry and fluid rheology. Indeed, for certain contraction ratios ($\beta = w_{0} / w_{c}$, the length scale ratio of uniform channel $w_{0}$ to the contraction width $w_{c}$) a lip vortex may form for $Wi < Wi_c$ (\cite{giesekus1968non}), as summarized by \cite{rothstein2001axisymmetric}. \cite{rothstein2001axisymmetric} also showed that the appearance of lip vortices in contraction flow is accompanied by a greater contribution of shear flow compared to extension-dominated flow where corner vortices manifest. More recently, viscoelastic contraction flow has received attention at the microscale, whereby inertia can be neglected and elastic effects are dominant (see the thorough review provided by \cite{rodd2007role}). For the negligible $Re$ regime, there is an apparent void of knowledge regarding three-dimensional (3D) flow at moderate to high $Wi$. Due to the difficulty of resolving 3D flows at the microscale, and the immense computational burden of solving a transient 3D elastic flow numerically, this flow type has not been fully detailed either experimentally or numerically. 
\begin{figure}
\begin{center}
\includegraphics[width=0.74\textwidth]{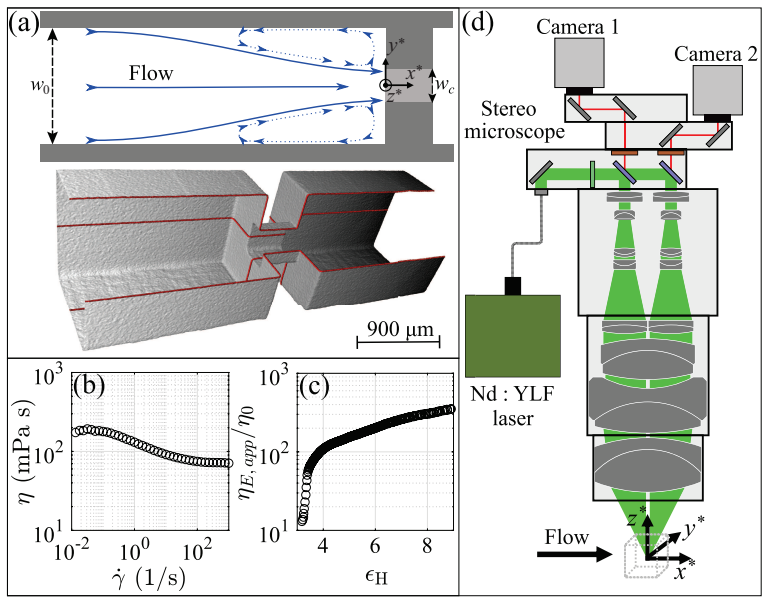}
    \caption{(a) A sketch and micro-CT scan of the glass square-sectioned contraction-expansion channel. (b, c) Shear and extensional rheology of the polyacrylamide test solution. (d) A diagram of the microtomographic PIV apparatus.}\label{fig01}
  \end{center}
\end{figure}

To date, elastic contraction flow has been studied primarily via localized or planar measurements such as particle image velocimetry (PIV), laser doppler velocimetry (LDV), or streak imagery. Global pressure measurements have also been employed to provide insights into drag reduction. In recent years, tomographic PIV (TPIV) has received increasing attention as a method whereby 3D flow volumes can be resolved via the reconstruction of a particle-laden flow from overlapping lines of sight, followed by cross-correlation between subsequent particle volumes (\cite{elsinga2006tomographic}). This method can also be applied at the microscale ($\upmu$-TPIV), with multiple lines of sight provided by stereomicroscopy.  Holographic PIV (HPIV) has also shown success in taking volumetric viscoelastic flow measurements in microscale geometries (\cite{qin2019upstream, qin2020three}), reporting a bistable negative wake ahead of a cylinder and out-of-plane instability modes along the flow separatrix of a cross-channel. However, HPIV is quite limited in terms of volume depth and reconstruction resolution compared to $\upmu$-TPIV (\cite{schafer2011comparison}). Nonetheless, the novel results from Qin \textit{et al}. (2019, 2020) suggest that, despite decades of research on fundamental viscoelastic flows, deep insights are still yet to be elucidated once out-of-plane dynamics are captured. 

Here, using a dilute solution of a high molecular weight polymer, we report the first investigations of 3D viscoelastic contraction flow at the microscale using the $\upmu$-TPIV method. We focus on the range of $Wi$ encompassing the transition from the vortex growth regime (which is accompanied by the growth of a steady central jet) to the onset of periodic vortex procession (which is accompanied by the circulation of the jet). We demonstrate that the circulation of the jet has phase-wise asymmetry dependence on the nominal $Wi$. By fully resolving the 3D velocity field, we can assess the true velocity gradient tensor and thus the rate-of-strain tensor. We show that the procession of the corner vortex is driven by the central jet continuously retreating from regions of increased rate-of-strain, and hypothesize that the underlying driving mechanism is localized strain-hardening of the polymer solution. 

\section{Experimental set-up}
\label{sec:experiments}

\subsection{Flow cell and viscoelastic fluid}

The experiments were conducted in a square-sectioned contraction-expansion flow cell (figure \ref{fig01}(a)) fabricated from fused-silica glass via selective laser-induced etching (\cite{gottmann2012digital}) using a commercial LightFab 3D printer (LightFab GmbH). This process can resolve features on the micron scale, with a surface r.m.s. of approximately 1 $\mathrm{\upmu m}$ (\cite{pimenta_etal_2020}). We measured the channel width and height from an x-ray microtomography scan (figure~\ref{fig01}(a)) as $w_0 = 860\pm10\; \mathrm{\upmu m}$ outside the contraction, and $w_c = 255\pm5\; \mathrm{\upmu m}$ inside the contraction, yielding a contraction ratio $\beta = 3.4$. Figure~\ref{fig01}(a) displays our dimensionless coordinate system, where each component is reduced by $w_c/\mathrm{2}$ (e.g., $x^* = 2x/w_c$). Flow was stepped to each velocity by two syringe pumps in a push-pull configuration. 

The viscoelastic test fluid was a polymeric solution composed of 107 parts-per-million (ppm) partially hydrolyzed polyacrylamide (HPAA, $M_W$ = 18 MDa, Polysciences Inc., U.S.A.), in a solvent of 85~wt\% glycerol and 15~wt\% deionized water. The refractive index of the fluid is closely matched to that of the fused-silica flow cell. An Anton-Paar MCR 502 stress-controlled rheometer was used with a cone and plate geometry (50 mm diameter, 1$^{\circ}$ angle) to characterize the shear viscosity of the fluid under steady shear. Figure~\ref{fig01}(b) shows that the fluid is weakly shear thinning and has a zero shear viscosity of $\eta_0$ = 184 mPa$\cdot$s. The relaxation time of the fluid was measured as $\lambda = 0.65$~s using a capillary breakup extensional rheometer (Haake CaBER, Thermo Fisher Scientific, see \cite{anna2001elasto}) fitted with $d_0 = 6$~mm diameter endplates. We plot the ratio of the apparent extensional viscosity $\eta_{E, app}$ to the zero shear viscosity $\eta_{0}$, against the accumulated Hencky strain $\epsilon_H = 2\mathrm{ln}(d_{0}/d(t))$ in figure~\ref{fig01}(c). The fluid exhibits strong strain-hardening with $\eta_{E,app} \approx 400\eta_{0}$ at high strains. The zero shear viscosity $\eta_{0}$ and the characteristic length scale $w_c/2$ are used to calculate $Re = \rho u_{c}w_{c}/2\eta_{0}$ and $Wi = 2\lambda u_{c}/w_{c}$. In the present work, $Re \lesssim 10^{-2}$ and as such is considered negligible. 

\subsection{Microtomographic PIV ($\upmu$-TPIV)}

Volumetric flow measurement can be achieved by the TPIV method, which is termed $\upmu$-TPIV when conducted via stereomicroscopy. As implemented in a LaVision FlowMaster system (LaVision GmbH), $\upmu$-TPIV uses a stereomicroscope (SteREO V20, Zeiss AG, Germany) with dual high speed cameras (Phantom VEO 410, 1280 x 800 pixels) imaging a fluid volume illuminated by a coaxial Nd:YLF laser (dual-pulsed, 527 nm wavelength), see figure~\ref{fig01}(d). The fluid was seeded with $2 \upmu$m diameter fluorescent particles (PS-FluoRed, Microparticles GmbH, Germany) to a visual concentration of 0.04 particles-per-pixel.

The flow was recorded as double-frame images captured at 12~Hz, with a flow-rate dependent time interval between laser pulses $\Delta t$ such that no particle moved more than 8 pixels. Frames were pre-processed with local background subtraction and Gaussian smoothing at $3 \times 3$~pixels. 3D calibration was performed by capturing reference images of a micro-grid at the planes $z = \pm 450~\upmu$m and $z= 0~\upmu$m, fully encompassing the depth of the flow cell ($w_0$), and a coordinate system was interpolated between these planes using a third-order polynomial. Particle positions in 3D were reconstructed from the images using four iterations of the Fast MART (Multiplicative Algebraic Reconstruction Technique) algorithm implemented in the commercial PIV software DaVis 10.1.2 (Lavision GmbH). Fast MART initializes the particle volume using the multiplicative line-of-sight routine (MLOS \cite{worth_nickels_2008, atkinson_soria_2009}), followed by iterations of Sequential MART (SMART \cite{atkinson_soria_2009}). We concluded the algorithm with five iterations of the Motion Tracking Enhancement (MTE) method (\cite{novara_etal_2010, lynch_scarano_2015}) to reduce spurious "ghost" particles which arise from randomly overlapping lines of sight (\cite{elsinga2006experimental}), and thus do not correlate in time. Volume self-calibration (\cite{wieneke_2008}) was employed to improve the accuracy of reconstruction. Particle displacements between particle volumes were obtained using a multi-grid iterative cross-correlation technique, with the final pass at $32 \times 32 \times 32$ voxels with 75\% overlap for a vector grid of $31~\upmu$m. To reduce measurement noise the vector field was spatiotemporally filtered with a second-order polynomial regression across neighborhoods of $5^3$ vectors in space and extended through five increments in time for a total kernel size of $5^4$ points. This polynomial regression visibly reduced measurement noise while predominantly preserving space-time resolution owing to the small kernel size: the filtering wavelength is substantially smaller than the flow dynamics reported in this work. This is a common approach to de-noise TPIV data (\cite{scarano2009three, elsinga2010three, schneiders2017resolving}). Ultimately we resolved 1600 flow volumes per recording: a duration of 133~s ($200\lambda$).

Uncertainty quantification for TPIV is a topic of ongoing research (\cite{atkinson2011accuracy,sciacchitano2019uncertainty}), but \textit{a priori} comparisons of experimental velocity fields to direct numerical solutions or synthetic reconstructions have yielded a TPIV uncertainty on the order of 0.1--0.3 pixels (\cite{atkinson2011accuracy}). As an \textit{a posteriori} assessment of our measurement quality, we validated conservation of mass for time-averaged flow volumes (\cite{zhang1997turbulent}). Flow divergence $\bnabla\bcdot\pmb{u}$ was calculated at each vector, and the error assessed relative to an assumption of incompressible flow (relative error $\zeta = (\bnabla\bcdot\pmb{u})^2 / tr(\bnabla\pmb{u}\bcdot\bnabla\pmb{u})$). The value $\zeta$ averaged 0.25 for $3\leq Wi \leq 87$. Divergence error relative to the magnitude of vorticity was 0.14 at $Wi=87$, in good agreement with a value of 0.2 from a three-camera TPIV experiment by \cite{kempaiah20203}.

\section{Results and discussion}
\label{sec:results}
 
Measurements were taken over a range of flow velocities encompassing $3\leq Wi \leq 87$ for a region of interest upstream of the contraction. Note that the downstream (expansion) side was imaged in a separate series of experiments, but flow remained steady across the $Wi$ range investigated. Throughout the discussion of the flow field kinematics we nondimensionalize lengths by $w_c/2$ and velocities by $U_c$ (the average flow velocity in the contraction). Deformation rates are hence reduced by $2U_c/w_c$. Times are nondimensionalized either by the fluid relaxation time $\lambda$ or by the fundamental period of circulation in the case of periodic flows. All nondimensional quantities are indicated by a superscript `*'. 

\subsection{Steady flow at low $Wi$}
\begin{figure}
\begin{center}
\includegraphics[width=0.74\textwidth]{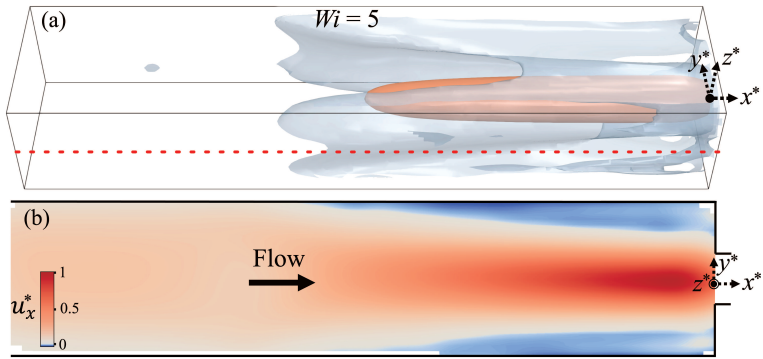}
    \caption{Averaged (a) isosurfaces and (b) midplane $x^*$\nobreakdash-$y^*$ slice of the streamwise velocity $u_x^*$ at $Wi = 5$. The isosurfaces in (a) are pink for $u_x^* =1/3$, and grey for $u_x^* = 0$. The dashed red line in (a) marks the $u_x^*$ extraction used in figure~\ref{fig03}.}
    \label{fig02}
  \end{center}
\end{figure}

 As flow approaches the contraction at $Wi\leq 5$, a steady separation point forms where the flow separates from the walls of the channel and the flow passes as central jet through the constriction. Figure \ref{fig02} presents (a) average isosurfaces and (b) a midplane $x^*$\nobreakdash-$y^*$ slice of the streamwise velocity $u_x^*$ for $Wi = 5$. The pink isosurface shows $u_x^* = 1/3$ (i.e., the central jet), while the grey surface marks $u_x^* = 0$ (i.e., the edges of the recirculating regions). Flow separation follows the upstream downzero crossing of $u_x^*$; the central jet is characterized by positive $u_x^*$, while the corner vortices drive negative $u_x^*$ backflow along the walls towards the separation point. We did not observe any lip vortices here, which agrees with indications from the literature whereby lip vortices are unlikely for abrupt contractions with $\beta > 2$ (\cite{rothstein2001axisymmetric}), signaling that extensional flow is dominant over shear flow.   

\subsection{The onset of periodic instability}
\begin{figure}
\begin{center}
\includegraphics[width=0.94\textwidth]{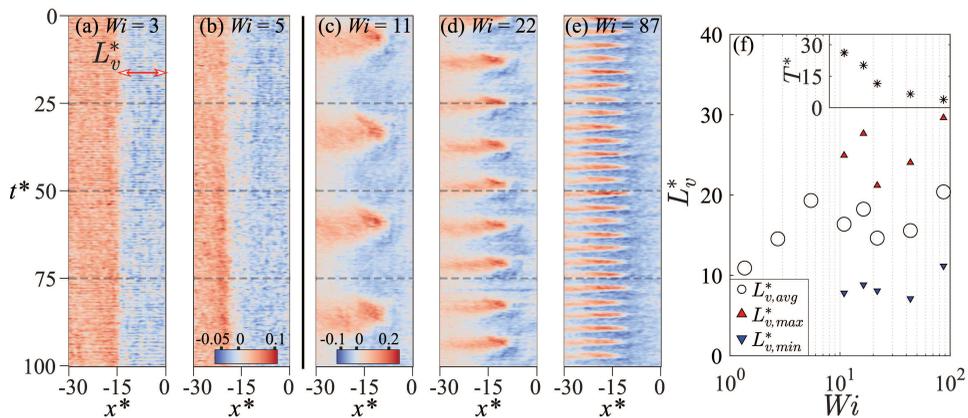}
    \caption{(a-e) Space-time diagrams of the streamwise velocity $u_x^*$ (indicated by the colour bars) along the wall at the $x^*$\nobreakdash-$y^*$ midplane (represented as a dashed line in figure~\ref{fig02}(a)). (f) The mean and range of values of $L^*_{v}$, with the fundamental period $T^*$ in the insert.}

    \label{fig03}
  \end{center}
\end{figure}

For increasing $Wi$, the corner vortex propagates upstream but remains steady, until, for $Wi>Wi_p$, the flow transitions to a periodic instability characterised by axial fluctuation of the upstream separation point and a circumferential procession of the central jet (see movie 1 in the supplemental material for animations of the jet for $5 \leq Wi \leq 44$). Figure \ref{fig03}(a-e) present space-time diagrams depicting the streamwise flow velocity $u_x^*(x^*)$ along the line $(y^*, z^*)=(-w_0/w_c, 0)$ (dashed red line in figure \ref{fig02}(a)) for five representative values of $3\leq Wi \leq 87$. 
The dimensionless vortex length $L^*_v$ is determined by the zero-crossing of $u_x^*$ (as indicated on figure \ref{fig03}(a)). The average value of  $L^*_v$ ($L_{v,avg}^*$) and the range of oscillation between  $L_{v,max}^*$ and  $L_{v,min}^*$ are plotted as a function of $Wi$ in figure \ref{fig03}(f), indicating a transition from steady vortex growth at lower $Wi$ to a regime of oscillation whereby $L_v^*$ apparently no longer scales directly with $Wi$. The inset of figure \ref{fig03}(f) shows the period of oscillation ($T^* = T / \lambda$) determined by FFT of the $L_v^*(t)$ signals.  
Above the critical value $5< Wi_p \leq 11$ for the onset of oscillation, the range of oscillation reaches a local maximum by $Wi=16$, and the mean separation point actually moves downstream (i.e., $L_v^*$ reduces) for $22\leq Wi \leq 44$. Interestingly, the reduction in $L_v^*$ does not affect the frequency of the instability, with the period of oscillation continuously decreasing from $T^*= 26$ at $Wi = 11$ to $T^*= 7$ at $Wi = 44$ (figure~\ref{fig03}(f)). Our results expand on prior experimental observations made in axisymmetric abrupt contractions which saw the vortex size increase monotonically for increasing $Wi$, although at a lower $Wi$ range than probed here due to larger length scales involved (e.g., $Wi < 5$ for \cite{mckinley1991nonlinear} and $Wi < 8$, for \cite{rothstein1999extensional}). In planar microfluidic contraction flow experiments, \cite{rodd2007role} reported a decrease in $L^*_v$ for a single data point at the maximum $Wi = 24$ acheived, but were unable to extrapolate the trend. Numerical work by \cite{comminal2016vortex} observed an $L^*_v$ plateau accompanied by periodic vortex annihilation for $Wi > 14$ in a 2D contraction, which they attributed to an accumulation of elastic strain upstream of the contraction. However, they acknowledged that their use of the Oldroyd-B constitutive model (\cite{oldroyd1950formulation}) lacks physical mechanisms (such as finite-extensibility) which would otherwise limit elastic stress.

\subsection{Out of plane jet dynamics}
As shown in the $x^*$\nobreakdash-$t^*$ space-time diagrams in figure \ref{fig03}(a-e), the flow instability is strongly periodic for $Wi \geq 11$. Thus, to collapse the dataset and further reduce noise we deploy time synchronous averaging (TSA) to reduce the time dimension into a single average cycle. This method is further discussed in \cite{bechhoefer2009review}. We isolated the time series of velocity magnitude at a single point in the volume, then used the local maxima in the time series at that point to segregate each period at all points in the volume. The cycles are mapped to a normalized phase time $\phi^*$ ranging from 0 to 1 (i.e., 1 cycle), and averaged into $f_{s}T$ bins where $f_{s}$ is the sampling frequency (12~Hz) and $T = T^*/\lambda$ is the average period ($T^*$ shown in figure \ref{fig03}(f)). Henceforth, phase averaged quantities will be marked by $\langle\ \rangle$.

\begin{figure}
\begin{center}
\includegraphics[width=0.82\textwidth]{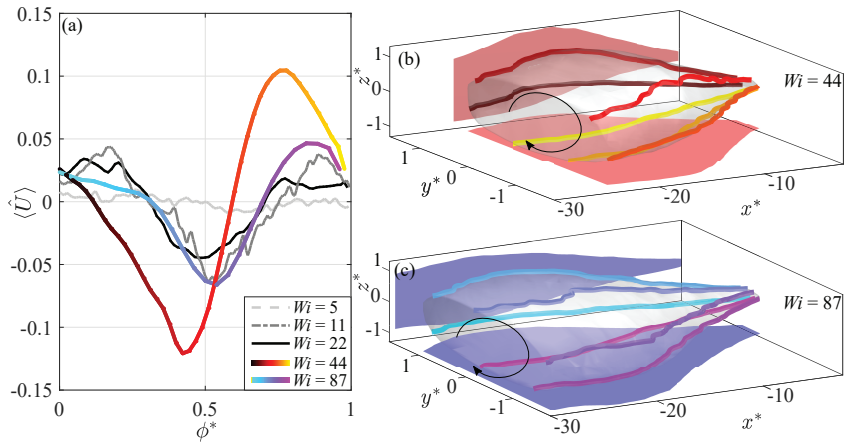}
    \caption{(a) The axially-averaged fluctuating velocity of the jet over phase time $\phi^*$. (b, c) Trajectories (coloured by $\phi^*$ in (a)) and planar projections of the jet at $Wi$ = 44 and 87.}

    \label{fig04}
  \end{center}
\end{figure}

We used the phase-binned data to investigate the 3D trajectory of the central jet passing through the toroidal corner vortex. We found the maximum flow velocity in each $y^*$\nobreakdash-$z^*$ plane along the $x^*$ axis for $-L_v^* \leq x^* \leq 0$. Thus, we extract $x^*$\nobreakdash-$y^*$\nobreakdash-$z^*$ trajectories of the circulating inner jet as it approaches the contraction. To quantify fluctuations in the jet velocity throughout phase time, we calculate the normalized fluctuating velocity along $x^*$ as $\langle \hat{U} \rangle(\phi^*) = (1/n_{x^*})\sum_{x^*=0}^{x^*=L_v^*}(\langle U \rangle^*(x^*,\phi^*)-\bar{\langle U\rangle^*}(x^*))/\langle U \rangle^*(x^*)_{max}$. In this way $\langle \hat{U} \rangle$ is not affected by the accelerating flow velocity as the jet approaches the contraction. Data of $\langle \hat{U} \rangle$ are presented in figure \ref{fig04}(a) for $Wi \geq 5$, along with snapshots of the location of the jet in figure \ref{fig04}(b, c) for $Wi = 44$ and $Wi = 87$. We observe a phase-wise asymmetry of the jet in both the fluctuating component of velocity and the jet location. For each $Wi$, the jet starts at the same place in the channel at $\phi^* =0$, but with an opposite directionality about $x^*$ for $Wi < 16$ (the direction of circulation is seemingly random between experiments). Maxima and minima in $\langle \hat{U} \rangle$ are evident for $Wi = 11$ near $\phi^*$ = 0.18, 0.88 and 0.56, respectively, and apparent but slightly degraded by $Wi=22$. The fact that the velocity fluctuation is small ($<5\%$) and almost symmetric about $\phi^* = 0.5$ indicates that the cyclical fluctuation in the $11\leq Wi \leq 22$ regime is likely geometric in origin as the channel has an $\approx 5~\upmu$m (or $\approx 0.02w_c$) variation between the width and height. At higher $Wi$ (e.g., $Wi = 44$ and $Wi=87$), the jet had the same directionality as for $Wi = 22$ and was mapped to the same locations in phase time, but strikingly the dual $\langle \hat{U} \rangle$ peaks are eliminated. Instead, we observe a single minimum and maximum for $\langle \hat{U} \rangle$ and the values occur at different phase times from those seen at lower velocities. Thus it is inferred that the peak fluctuation is independent of the $y^*$\nobreakdash-$z^*$ coordinate between different $Wi$. To relate velocity fluctuations within the jet to the spatial distribution, we present in figure \ref{fig04}(b, c) trajectories of the jet core position in $x^*$\nobreakdash-$y^*$\nobreakdash-$z^*$ over phase time. Here we see the minimum $\langle \hat{U} \rangle$ for $Wi=44$ coincides with a skewed curvature of the jet towards $+z^*$ near $\phi^* = 0.42$, and the jet is comparatively tighter towards $z^* = 0$ which aligns in time with the maximum $\langle \hat{U} \rangle$ near $\phi^* = 0.78$: the distribution of the jet varies inversely with the fluctuating velocity such that flow rate is preserved. A similar effect is seen for $Wi= 87$, but with a shifted phase timing of $\langle \hat{U} \rangle$, placing the asymmetry more in the $x^*$\nobreakdash-$y^*$ plane. Reconfiguration to single-sided asymmetry in the jet at $Wi\geq44$ coincides with a phenomenon previously discussed: the growth plateau in $L^*_v$ from $11\leq Wi \leq 22$ despite a decreasing core period of the circulating jet (figure \ref{fig03}(f)). Whereas the symmetrically swirling jet appears to prohibit vortex growth, a break to strong asymmetry above $Wi = 22$ proves a preferable route and $L^*_v$ again increases for increasing $Wi$. 

\subsection{The role of the rate-of-strain tensor}

Dilute solutions of high molecular weight polymers are known to shear-thin under shear flow, and strain-harden under extensional deformations (\cite{tirtaatmadja1993filament, solomon1996transient}), as we observed for the HPAA fluid used in our work (figure~\ref{fig01}(b, c)). Furthermore, as summarized in \cite{rothstein2001axisymmetric}, the lack of lip vortices observed in our experiments implies that extensional flow is dominant over shear flow. We can gain qualitative insight into the relevance of strain-hardening on the dynamics of our micro-contraction flow by considering the local rate-of-strain tensor $\pmb{D} = (\bnabla \langle\pmb{u}\rangle+\bnabla \langle\pmb{u}\rangle^\intercal)/2$. Here we rely solely on the $\upmu$-TPIV measurements to obtain the velocity vector field before calculating $\pmb{D}$. We take the magnitude of $\pmb{D}$ as $\langle\dot{\gamma}\rangle = \sqrt{2(\pmb{D}:\pmb{D})}$ (reduced as $\langle\dot{\gamma}\rangle^* = \langle\dot{\gamma}\rangle w_c / (2 U_c)$) to highlight regions of high rate-of-strain where strain-hardening of the HPAA is more likely. 
\begin{figure}
\begin{center}
\includegraphics[width=0.99\textwidth]{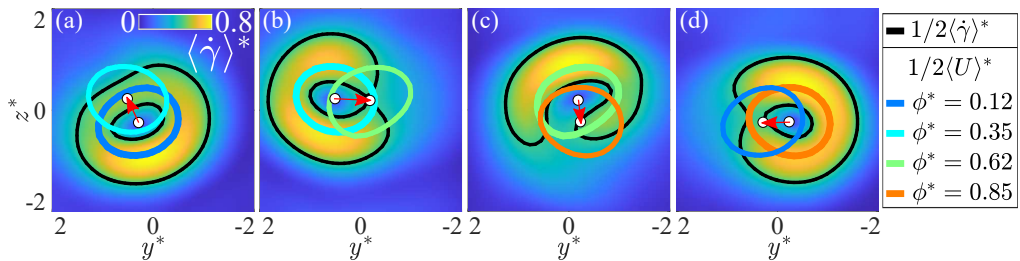}
    \caption{ $y^*$\nobreakdash-$z^*$ slices of $\langle\dot{\gamma}\rangle^*$ at $x^* = -8$, with black contours of $0.5\langle\dot{\gamma}\rangle^*$ and coloured contours of $0.5\langle U\rangle^*$ for $Wi = 87$, for increasing normalized phase time $\phi^*$. }
    \label{fig05}
  \end{center}
\end{figure}
We extracted $y^*$\nobreakdash-$z^*$ slices at an arbitrary $x^* = -8$ for $Wi = 87$ to show a simplified perspective on relationship between $0.5\langle\dot{\gamma}\rangle^*$ and the location of the jet (coloured contours of $0.5\langle U \rangle^*$, $\langle U \rangle^*= \langle U \rangle / U_{c}$) in figure \ref{fig05}(a-d). The subplots progress along phase time $\phi^*$, with each plane including the $0.5\langle U \rangle^*$ contour from the following phase time. Two trends are noted: first that while the jet (the solid coloured contour) stays centered about $\langle\dot{\gamma}\rangle^*$, the jet location forward in time is always towards the exterior of the $\langle\dot{\gamma}\rangle$ contour. Secondly, a phase-wise asymmetry manifests in the distribution of $\langle\dot{\gamma}\rangle^*$. $\phi^* = 0.12$ has low asymmetry in $\langle \hat{U} \rangle$ (figure \ref{fig05}(b)), as well as in the distribution of $\langle\dot{\gamma}\rangle$. By $\phi^* = 0.62$, $\langle \hat{U} \rangle$ is highly deviated and this aligns with a strong mismatch in $\langle\dot{\gamma}\rangle^*$ about the circumference of the jet. Therefore, it appears that a mismatch in rate-of-strain about the jet can strongly influence the phasewise progression of the jet as it circulates, with positive and negative fluctuations in $\langle \hat{U} \rangle$ accompanied by an unbalanced distribution of  $\langle\dot{\gamma}\rangle^*$. 

\begin{figure}
\begin{center}
\includegraphics[width=0.95\textwidth]{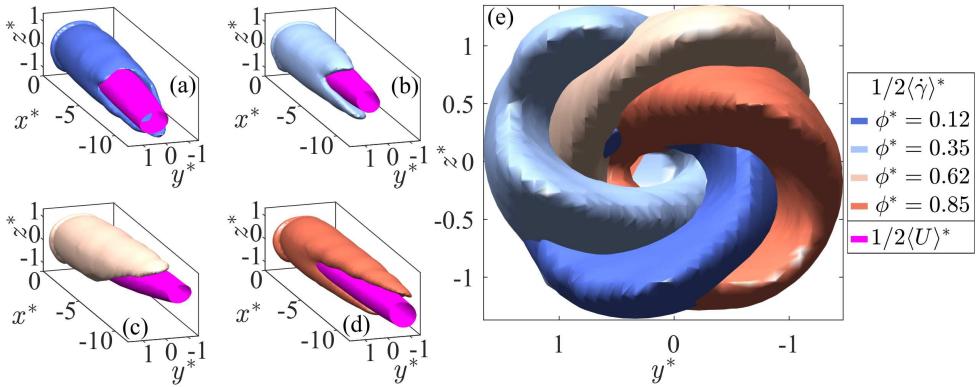}
    \caption{(a-d) Phase-averaged isosurfaces of $0.5\langle\dot{\gamma}\rangle^*$ and $0.5\langle U\rangle^*$ for $Wi= 87$. (e) The $-x^*$ projection of the $\langle\dot{\gamma}\rangle^*$ surfaces from (a-d), coloured by their normalized phase time $\phi^*$. For advancing time $\phi^*$, the inner jet displaces towards decreasing rate-of-strain.}
    \label{fig06}
  \end{center}
\end{figure}

In figure \ref{fig06}, we present isosurfaces of $0.5\langle\dot{\gamma}\rangle^*$ and $0.5\langle U \rangle^* $, the maximum rate-of-strain and flow velocity, for $Wi= 87$. Figure \ref{fig06}(a-d) show 4 timesteps throughout a circulation, where the volume of high rate-of-strain forms a band about the circumference of the core of the jet (the pink isosurface). Moreover, the $\langle\dot{\gamma}\rangle^*$ isosurface extends further upstream on one side of the jet for all timesteps, i.e., the rate-of-strain is greater along one side of the jet. An animated loop of the jet circulating with the rate-of-strain volume is shown in supplemental movie 2. 

We directly compare the rate-of-strain forward in time in figure \ref{fig06}(e) as projections of $\langle\dot{\gamma}\rangle^*$ from (a-d) from the $-x^*$ direction. Moving clockwise from the $\phi^*=0.12$ isosurface, each surface forward in time presents a decrease in the rate-of-strain in the clockwise direction. In other words, a perpetual retreat of the central jet from regions of increased rate-of-strain. The dynamics of elastic contraction flow has been reported to be sensitive to the extensional rheology (\cite{rothstein2001axisymmetric}), and since our HPAA solution strain-hardens (figure ~\ref{fig01}(c)), we can infer from the flow kinematics that the central jet moves about the contraction driven by gradients of strain-hardening HPAA, which will tend to follow the regions of increased rate-of-strain. This sheds new light onto the likely significance of extensional rheology for \textit{local} dynamics in viscoelastic contraction flow, as experimentally describing flow topology via directly resolving the rate-of-strain tensor has not been achieved hitherto for elastic flow instability. 

\section{Conclusions}
Using $\upmu$-TPIV, we have experimentally resolved for the first time the highly 3D dynamics of viscoelastic flow through a square-square micro-contraction at low $Re$ and high $Wi$. We captured steady and periodic flow instability for a central jet of fluid passing through a toroidal vortex pinned about the contraction entrance, and observed a new vortex growth plateau whereby the period of instability decreases with $Wi$ but the enveloped vortex volume stagnates. For low $11\leq Wi\leq 22$, the jet circulates with a symmetric cyclical velocity fluctuation likely originating from geometric imperfections. This region coincides with the vortex growth plateau. At higher $Wi\geq44$, a strong asymmetry in the jet forms which corresponds to exiting the growth plateau. This would indicate that the asymmetric mode provides a preferable route to vortex growth. We determined the first experimental mapping of the full rate-of-strain tensor to transient dynamics for viscoelastic flow instabilities. We relate regions of increased rate-of-strain and the extensional rheology of the fluid to the directionality of the circulating jet. Gradients of strain-hardening in the fluid provide a likely circulation mechanism as the jet retreats from locally strain hardened regions of the flowing viscoelastic polymer solution: a new insight on the significance of extensional rheology to local dynamics of viscoelastic contraction flow. 




\backsection[Funding]{D.W.C., A.Q.S, and S.J.H. acknowledge financial support from the Japanese Society for the Promotion of Science (JSPS, Grant Nos. 21K14080, 18K03958, 18H01135, and 21K03884).}

\backsection[Declaration of interests]{The authors report no conflict of interest.}


\begin{thebibliography}{34}
\expandafter\ifx\csname natexlab\endcsname\relax\def\natexlab#1{#1}\fi
\def\au#1{#1} \def\ed#1{#1} \def\yr#1{#1}\def\at#1{#1}\def\jt#1{\textit{#1}}
  \def\bt#1{#1}\def\bvol#1{\textbf{#1}} \def\vol#1{#1} \def\pg#1{#1}
  \def\publ#1{#1}\def\arxiv#1{#1}\def\org#1{#1}\def\st#1{\textit{#1}}

\bibitem[Afonso {\em et~al.\/}(2011)Afonso, Oliveira, Pinho \&
  Alves]{afonso2011dynamics}
{\sc \au{Afonso, A.M.}, \au{Oliveira, P.J.}, \au{Pinho, F.T.} \& \au{Alves,
  M.A.}} \yr{2011}  \at{Dynamics of high-{D}eborah-number entry flows: a
  numerical study}.  \jt{Journal of Fluid Mechanics}  \bvol{677},  \pg{272}.

\bibitem[Anna \& McKinley(2001)]{anna2001elasto}
{\sc \au{Anna, S.L.} \& \au{McKinley, G.H.}} \yr{2001}  \at{Elasto-capillary
  thinning and breakup of model elastic liquids}.  \jt{Journal of Rheology}
  \bvol{45}~(1),  \pg{115--138}.

\bibitem[Atkinson {\em et~al.\/}(2011)Atkinson, Coudert, Foucaut, Stanislas \&
  Soria]{atkinson2011accuracy}
{\sc \au{Atkinson, C.}, \au{Coudert, S.}, \au{Foucaut, J-M}, \au{Stanislas, M.}
  \& \au{Soria, J.}} \yr{2011}  \at{The accuracy of tomographic particle image
  velocimetry for measurements of a turbulent boundary layer}.  \jt{Experiments
  in Fluids}  \bvol{50}~(4),  \pg{1031--1056}.

\bibitem[Atkinson \& Soria(2009)]{atkinson_soria_2009}
{\sc \au{Atkinson, C.} \& \au{Soria, J.}} \yr{2009}  \at{An efficient
  simultaneous reconstruction technique for tomographic particle image
  velocimetry}.  \jt{Experiments in Fluids}  \bvol{47}~(4-5),  \pg{553}.

\bibitem[Bechhoefer \& Kingsley(2009)]{bechhoefer2009review}
{\sc \au{Bechhoefer, E.} \& \au{Kingsley, M.}} \yr{2009} A review of time
  synchronous average algorithms.  \bt{In {\em Annual Conference of the
  Prognostics and Health Management Society\/}}, ,  \vol{vol.~23},  \pg{pp.
  1--10}.

\bibitem[Boger(1987)]{boger1987viscoelastic}
{\sc \au{Boger, D.V.}} \yr{1987}  \at{Viscoelastic flows through contractions}.
   \jt{Annual Review of Fluid Mechanics}  \bvol{19}~(1),  \pg{157--182}.

\bibitem[Comminal {\em et~al.\/}(2016)Comminal, Hattel, Alves \&
  Spangenberg]{comminal2016vortex}
{\sc \au{Comminal, R.}, \au{Hattel, J.H.}, \au{Alves, M.A.} \& \au{Spangenberg,
  J.}} \yr{2016}  \at{Vortex behavior of the {O}ldroyd-{B} fluid in the 4-1
  planar contraction simulated with the streamfunction--log-conformation
  formulation}.  \jt{Journal of Non-Newtonian Fluid Mechanics}  \bvol{237},
  \pg{1--15}.

\bibitem[Elsinga {\em et~al.\/}(2010)Elsinga, Adrian, Van~Oudheusden \&
  Scarano]{elsinga2010three}
{\sc \au{Elsinga, G.E.}, \au{Adrian, R.J.}, \au{Van~Oudheusden, B.W.} \&
  \au{Scarano, F.}} \yr{2010}  \at{Three-dimensional vortex organization in a
  high-{R}eynolds-number supersonic turbulent boundary layer}.  \jt{Journal of
  Fluid Mechanics}  \bvol{644},  \pg{35--60}.

\bibitem[Elsinga {\em et~al.\/}(2006{\natexlab{{\em a\/}}})Elsinga, Scarano,
  Wieneke \& van Oudheusden]{elsinga2006tomographic}
{\sc \au{Elsinga, G.E.}, \au{Scarano, F.}, \au{Wieneke, B.} \& \au{van
  Oudheusden, B.W.}} \yr{2006{\natexlab{{\em a\/}}}}  \at{Tomographic particle
  image velocimetry}.  \jt{Experiments in Fluids}  \bvol{41}~(6),
  \pg{933--947}.

\bibitem[Elsinga {\em et~al.\/}(2006{\natexlab{{\em b\/}}})Elsinga,
  Van~Oudheusden \& Scarano]{elsinga2006experimental}
{\sc \au{Elsinga, G.E.}, \au{Van~Oudheusden, B.W.} \& \au{Scarano, F.}}
  \yr{2006{\natexlab{{\em b\/}}}} Experimental assessment of tomographic-{PIV}
  accuracy.  \bt{In {\em 13th international symposium on applications of laser
  techniques to fluid mechanics, Lisbon, Portugal\/}}, ,  \vol{vol.~20}.

\bibitem[Giesekus(1968)]{giesekus1968non}
{\sc \au{Giesekus, H.W.}} \yr{1968}  \at{Non-linear effects in the flow of
  visco-elastic fluids through slits and holes}.  \jt{Rheological Acta}
  \bvol{7},  \pg{127--138}.

\bibitem[Gottmann {\em et~al.\/}(2012)Gottmann, Hermans \&
  Ortmann]{gottmann2012digital}
{\sc \au{Gottmann, J.}, \au{Hermans, M.} \& \au{Ortmann, J.}} \yr{2012}
  \at{Digital photonic production of micro structures in glass by in-volume
  selective laser-induced etching using a high speed micro scanner}.
  \jt{Physics Procedia}  \bvol{39},  \pg{534--541}.

\bibitem[Kempaiah {\em et~al.\/}(2020)Kempaiah, Scarano, Elsinga, van
  Oudheusden \& Bermel]{kempaiah20203}
{\sc \au{Kempaiah, K.U.}, \au{Scarano, F.}, \au{Elsinga, G.E.}, \au{van
  Oudheusden, B.W.} \& \au{Bermel, L.}} \yr{2020}  \at{3-dimensional particle
  image velocimetry based evaluation of turbulent skin-friction reduction by
  spanwise wall oscillation}.  \jt{Physics of Fluids}  \bvol{32}~(8),
  \pg{085111}.

\bibitem[Lynch \& Scarano(2015)]{lynch_scarano_2015}
{\sc \au{Lynch, K.P.} \& \au{Scarano, F.}} \yr{2015}  \at{An efficient and
  accurate approach to {MTE-MART} for time-resolved tomographic {PIV}}.
  \jt{Experiments in Fluids}  \bvol{56}~(3),  \pg{66}.

\bibitem[McKinley {\em et~al.\/}(1991)McKinley, Raiford, Brown \&
  Armstrong]{mckinley1991nonlinear}
{\sc \au{McKinley, G.H.}, \au{Raiford, W.P.}, \au{Brown, R.A.} \&
  \au{Armstrong, R.C.}} \yr{1991}  \at{Nonlinear dynamics of viscoelastic flow
  in axisymmetric abrupt contractions}.  \jt{Journal of Fluid Mechanics}
  \bvol{223},  \pg{411--456}.

\bibitem[Novara {\em et~al.\/}(2010)Novara, Batenburg \&
  Scarano]{novara_etal_2010}
{\sc \au{Novara, M.}, \au{Batenburg, K.J.} \& \au{Scarano, F.}} \yr{2010}
  \at{Motion tracking-enhanced {MART} for tomographic {PIV}}.  \jt{Measurement
  Science and Technology}  \bvol{21}~(3),  \pg{035401}.

\bibitem[Oldroyd(1950)]{oldroyd1950formulation}
{\sc \au{Oldroyd, J.G.}} \yr{1950}  \at{On the formulation of rheological
  equations of state}.  \jt{Proceedings of the Royal Society of London. Series
  A. Mathematical and Physical Sciences}  \bvol{200}~(1063),  \pg{523--541}.

\bibitem[Pimenta \& Alves(2017)]{pimenta2017stabilization}
{\sc \au{Pimenta, F.} \& \au{Alves, M.A.}} \yr{2017}  \at{Stabilization of an
  open-source finite-volume solver for viscoelastic fluid flows}.  \jt{Journal
  of Non-Newtonian Fluid Mechanics}  \bvol{239},  \pg{85--104}.

\bibitem[Pimenta {\em et~al.\/}(2020)Pimenta, Toda-Peters, Shen, Alves \&
  Haward]{pimenta_etal_2020}
{\sc \au{Pimenta, F.}, \au{Toda-Peters, K.}, \au{Shen, A.Q.}, \au{Alves, M.A.}
  \& \au{Haward, S.J.}} \yr{2020}  \at{Viscous flow through microfabricated
  axisymmetric contraction/expansion geometries}.  \jt{Experiments in Fluids}
  \bvol{61}~(9),  \pg{1--16}.

\bibitem[Qin {\em et~al.\/}(2020)Qin, Ran, Salipante, Hudson \&
  Arratia]{qin2020three}
{\sc \au{Qin, B.}, \au{Ran, R.}, \au{Salipante, P.F.}, \au{Hudson, S.} \&
  \au{Arratia, P.E.}} \yr{2020}  \at{Three-dimensional structures and symmetry
  breaking in viscoelastic cross-channel flow}.  \jt{Soft Matter}
  \bvol{16}~(30),  \pg{6969--6974}.

\bibitem[Qin {\em et~al.\/}(2019)Qin, Salipante, Hudson \&
  Arratia]{qin2019upstream}
{\sc \au{Qin, B.}, \au{Salipante, P.F.}, \au{Hudson, S.D.} \& \au{Arratia,
  P.E.}} \yr{2019}  \at{Upstream vortex and elastic wave in the viscoelastic
  flow around a confined cylinder}.  \jt{Journal of Fluid Mechanics}
  \bvol{864}.

\bibitem[Rodd {\em et~al.\/}(2007)Rodd, Cooper-White, Boger \&
  McKinley]{rodd2007role}
{\sc \au{Rodd, L.E.}, \au{Cooper-White, J.J.}, \au{Boger, D.V.} \&
  \au{McKinley, G.H.}} \yr{2007}  \at{Role of the elasticity number in the
  entry flow of dilute polymer solutions in micro-fabricated contraction
  geometries}.  \jt{Journal of Non-Newtonian Fluid Mechanics}
  \bvol{143}~(2-3),  \pg{170--191}.

\bibitem[Rothstein \& McKinley(1999)]{rothstein1999extensional}
{\sc \au{Rothstein, J.P.} \& \au{McKinley, G.H.}} \yr{1999}  \at{Extensional
  flow of a polystyrene {B}oger fluid through a 4: 1: 4 axisymmetric
  contraction/expansion}.  \jt{Journal of Non-Newtonian Fluid Mechanics}
  \bvol{86}~(1-2),  \pg{61--88}.

\bibitem[Rothstein \& McKinley(2001)]{rothstein2001axisymmetric}
{\sc \au{Rothstein, J.P.} \& \au{McKinley, G.H.}} \yr{2001}  \at{The
  axisymmetric contraction--expansion: the role of extensional rheology on
  vortex growth dynamics and the enhanced pressure drop}.  \jt{Journal of
  Non-Newtonian Fluid Mechanics}  \bvol{98}~(1),  \pg{33--63}.

\bibitem[Scarano \& Poelma(2009)]{scarano2009three}
{\sc \au{Scarano, F.} \& \au{Poelma, C.}} \yr{2009}  \at{Three-dimensional
  vorticity patterns of cylinder wakes}.  \jt{Experiments in Fluids}
  \bvol{47}~(1),  \pg{69--83}.

\bibitem[Sch{\"a}fer \& Schr{\"o}der(2011)]{schafer2011comparison}
{\sc \au{Sch{\"a}fer, L.} \& \au{Schr{\"o}der, W.}} \yr{2011} Comparison of
  holographic and tomographic particle-image velocimetry turbulent channel flow
  measurements.  \bt{In {\em Journal of Physics: Conference Series\/}}, ,
  \vol{vol. 318},  \pg{p. 022019}. IOP Publishing.

\bibitem[Schneiders {\em et~al.\/}(2017)Schneiders, Scarano \&
  Elsinga]{schneiders2017resolving}
{\sc \au{Schneiders, J.F.G.}, \au{Scarano, F.} \& \au{Elsinga, G.E.}} \yr{2017}
   \at{Resolving vorticity and dissipation in a turbulent boundary layer by
  tomographic {PTV} and {VIC+}}.  \jt{Experiments in Fluids}  \bvol{58}~(4),
  \pg{27}.

\bibitem[Sciacchitano(2019)]{sciacchitano2019uncertainty}
{\sc \au{Sciacchitano, A.}} \yr{2019}  \at{Uncertainty quantification in
  particle image velocimetry}.  \jt{Measurement Science and Technology}
  \bvol{30}~(9),  \pg{092001}.

\bibitem[Solomon \& Muller(1996)]{solomon1996transient}
{\sc \au{Solomon, M.J.} \& \au{Muller, S.J.}} \yr{1996}  \at{The transient
  extensional behavior of polystyrene-based {B}oger fluids of varying solvent
  quality and molecular weight}.  \jt{Journal of Rheology}  \bvol{40}~(5),
  \pg{837--856}.

\bibitem[Tirtaatmadja \& Sridhar(1993)]{tirtaatmadja1993filament}
{\sc \au{Tirtaatmadja, V.} \& \au{Sridhar, T.}} \yr{1993}  \at{A filament
  stretching device for measurement of extensional viscosity}.  \jt{Journal of
  Rheology}  \bvol{37}~(6),  \pg{1081--1102}.

\bibitem[White {\em et~al.\/}(1987)White, Gotsis \& Baird]{white1987review}
{\sc \au{White, S.A.}, \au{Gotsis, A.D.} \& \au{Baird, D.G.}} \yr{1987}
  \at{Review of the entry flow problem: experimental and numerical}.
  \jt{Journal of Non-Newtonian Fluid Mechanics}  \bvol{24}~(2),  \pg{121--160}.

\bibitem[Wieneke(2008)]{wieneke_2008}
{\sc \au{Wieneke, B.}} \yr{2008}  \at{Volume self-calibration for {3D} particle
  image velocimetry}.  \jt{Experiments in Fluids}  \bvol{45}~(4),
  \pg{549--556}.

\bibitem[Worth \& Nickels(2008)]{worth_nickels_2008}
{\sc \au{Worth, N.A.} \& \au{Nickels, T.B.}} \yr{2008}  \at{Acceleration of
  tomo-{PIV} by estimating the initial volume intensity distribution}.
  \jt{Experiments in Fluids}  \bvol{45}~(5),  \pg{847--856}.

\bibitem[Zhang {\em et~al.\/}(1997)Zhang, Tao \& Katz]{zhang1997turbulent}
{\sc \au{Zhang, J.}, \au{Tao, B.} \& \au{Katz, J.}} \yr{1997}  \at{Turbulent
  flow measurement in a square duct with hybrid holographic {PIV}}.
  \jt{Experiments in Fluids}  \bvol{23}~(5),  \pg{373--381}.

\end{thebibliography}

\end{document}